\documentclass[aps,twocolumn,showpacs,superscriptaddress,preprintnumbers,amsmath,amssymb]{revtex4}
\usepackage{graphicx}
\usepackage{dcolumn}
\usepackage{bm}
\usepackage{multirow}
\begin{document}
\preprint{preprint \today}
\title{Polarization-analyzed resonant inelastic x-ray scattering of the
orbital excitations in KCuF$_3$}
\author{K. Ishii}
\affiliation{SPring-8, Japan Atomic Energy Agency, Hyogo 679-5148,
Japan}
\author{S. Ishihara}
\affiliation{Department of Physics, Tohoku University, Sendai 980-8578,
Japan}
\affiliation{CREST, Japan Science and Technology Agency (JST), Tokyo
102-0075, Japan.}
\author{Y. Murakami}
\affiliation{SPring-8, Japan Atomic Energy Agency, Hyogo 679-5148,
Japan}
\affiliation{Department of Physics, Tohoku University, Sendai 980-8578,
Japan}
\affiliation{Institute of Materials Structure Science, Tsukuba 305-0801,
Japan}
\author{K. Ikeuchi}
\affiliation{SPring-8, Japan Atomic Energy Agency, Hyogo 679-5148,
Japan}
\affiliation{Institute of Materials Structure Science, Tsukuba 305-0801,
Japan}
\author{K. Kuzushita}
\affiliation{SPring-8, Japan Atomic Energy Agency, Hyogo 679-5148,
Japan}
\author{T. Inami}
\affiliation{SPring-8, Japan Atomic Energy Agency, Hyogo 679-5148,
Japan}
\author{K. Ohwada}
\affiliation{SPring-8, Japan Atomic Energy Agency, Hyogo 679-5148,
Japan}
\author{M. Yoshida}
\affiliation{SPring-8, Japan Atomic Energy Agency, Hyogo 679-5148,
Japan}
\author{I. Jarrige}
\affiliation{SPring-8, Japan Atomic Energy Agency, Hyogo 679-5148,
Japan}
\author{N. Tatami}
\affiliation{Department of Physics, Tohoku University, Sendai 980-8578,
Japan}
\author{S. Niioka}
\affiliation{Department of Physics, Tohoku University, Sendai 980-8578,
Japan}
\author{D. Bizen}
\affiliation{Department of Physics, Tohoku University, Sendai 980-8578,
Japan}
\author{Y. Ando}
\affiliation{Department of Physics, Tohoku University, Sendai 980-8578,
Japan}
\author{J. Mizuki}
\affiliation{SPring-8, Japan Atomic Energy Agency, Hyogo 679-5148,
Japan}
\author{S. Maekawa}
\affiliation{Advanced Science Research Center, Japan Atomic Energy
Agency, Tokai 319-1195, Japan.}
\affiliation{CREST, Japan Science and Technology Agency (JST), Tokyo
102-0075, Japan.}
\author{Y. Endoh}
\affiliation{SPring-8, Japan Atomic Energy Agency, Hyogo 679-5148,
Japan}
\affiliation{International Institute for Advanced Studies, Kizu, Kyoto
619-0025, Japan}
\date{\today}

\begin{abstract}
We report a Cu $K$-edge resonant inelastic x-ray scattering (RIXS) study
of orbital excitations in KCuF$_3$.  By performing the polarization
analysis of the scattered photons, we disclose that the excitation
between the $e_g$ orbitals and the excitations from $t_{2g}$ to $e_g$
exhibit distinct polarization dependence.  The polarization dependence
of the respective excitations is interpreted based on a phenomenological
consideration of the symmetry of the RIXS process that yields a
necessary condition for observing the excitations.  In addition, we show
that the orbital excitations are dispersionless within our experimental
resolution.
\end{abstract}


\maketitle

Strongly correlated transition metal compounds attract great interest
because of a variety of interesting properties such as high-temperature
superconductivity in cuprates and colossal magnetoresistance in
manganites \cite{Maekawa2004}.  It is widely recognized that the three
degrees of freedom of the $d$ electron, i.e., charge, spin, and orbital,
play a crucial role in the occurrence of these phenomena.  Hence
spectroscopic investigations to measure the excitations of these degrees
of freedom are required to elucidate their underlying interactions.  Since
intra-atomic $d$-$d$ excitations are forbidden within the dipole
approximation, spectroscopic methods other than the conventional optical
measurement are highly demanded.  In this respect, inelastic x-ray
scattering (IXS) has proven to be an ideal candidate.  In addition to
the capability of observing dipole-forbidden transitions, IXS has a
great advantage of also offering momentum resolution.  Especially, one
can explore a wide momentum range using IXS in the hard x-ray regime.  A
recent non-resonant IXS study showed that the $d$-$d$ excitations can be
observed at high momentum transfer \cite{Larson2007}, although
non-resonant IXS usually suffers from low intensity.  This shortcoming
can be overcome by using resonant inelastic x-ray scattering (RIXS),
where the incident photon energy is tuned near an absorption edge of a
constituent element, resulting in a significant resonant enhancement of
the electronic excitations \cite{Kotani2001,Hasan2000a}.  RIXS spectra
are primarily related to the dynamical charge correlation
\cite{Ishii2005b,JKim2009}.  Furthermore, spin
\cite{Hill2008,Braicovich2009,Schlappa2009a} and orbital
\cite{Kim2004a,Ulrich2009a} excitations have also become accessible with
the more recently developed spectrometers and their improved energy
resolution.

In general, the RIXS spectrum is a function of energy, momentum, and
polarization of both the incident and scattered photons. However, most
RIXS studies so far have focused on energy and momentum dependences
while the polarization, which is an inherent and important character of the
photon, was overlooked.  Even though the role of the incident photon
polarization was discussed in relation with the resonant conditions in a
few experimental and theoretical works
\cite{Hamalainen2000,Id'e2000,Lu2006,Takahashi2008,Vernay2008}, the
scattered photon polarization has not been identified at all.  Like
conventional Raman spectroscopy, polarization in RIXS must be connected
to the symmetry of the excitations.  Because the symmetry argument is
rigorous and independent of the parameters in theoretical models, the
polarization can be very useful for the assignment of the excitations in
RIXS.

\begin{figure*}[t]
\includegraphics[scale=0.3]{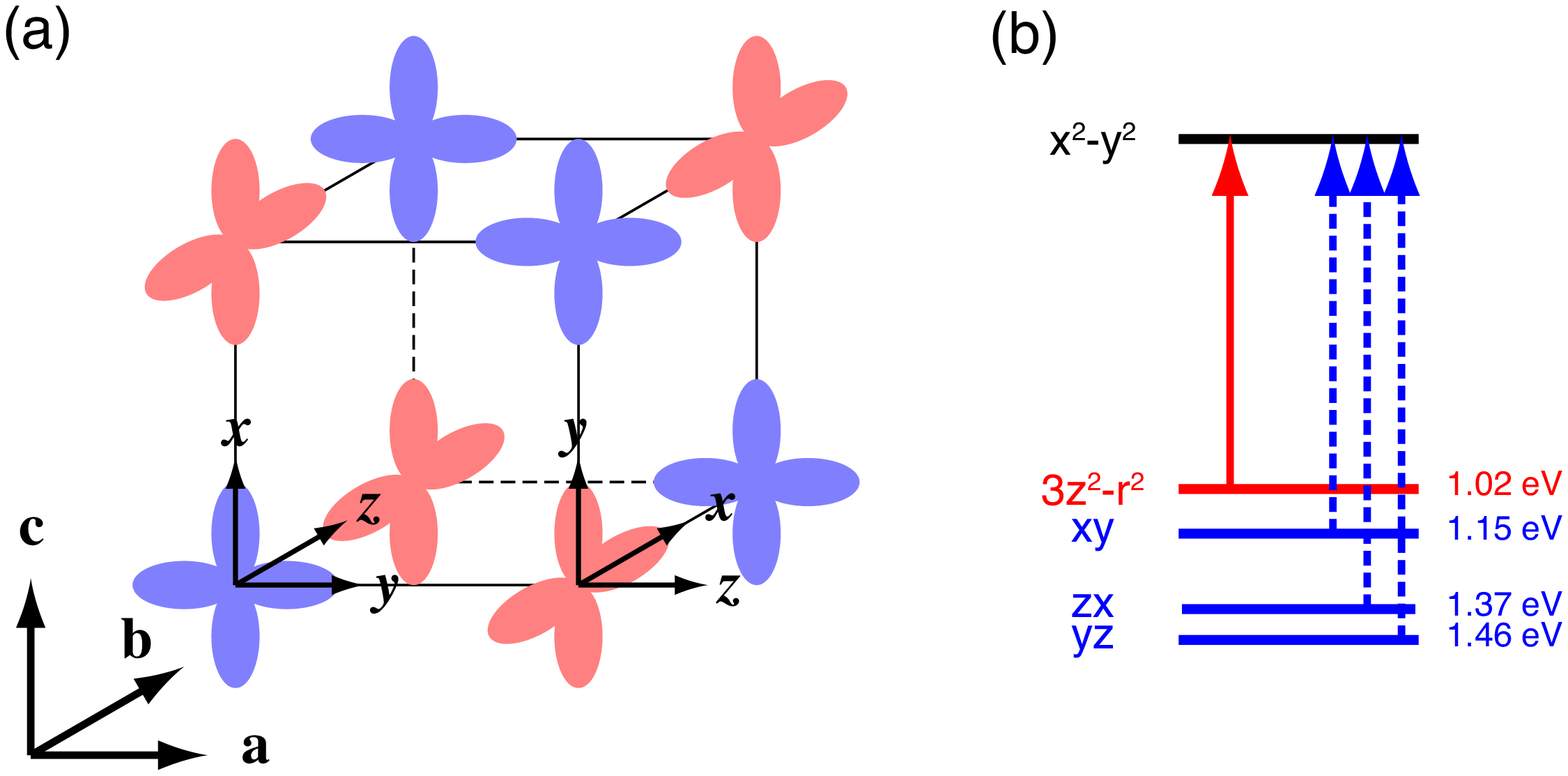}
\includegraphics[scale=0.3]{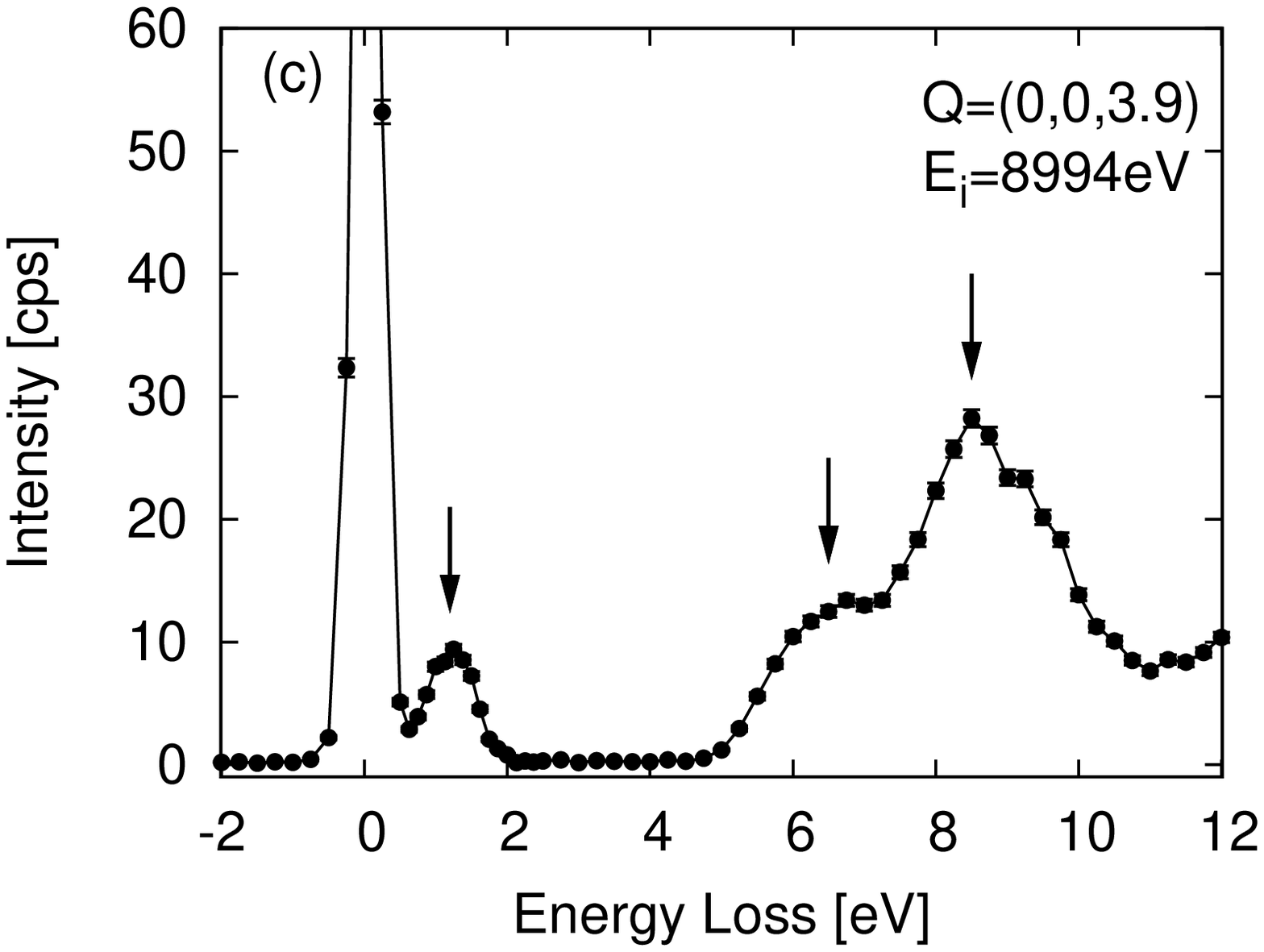}
\includegraphics[scale=0.3]{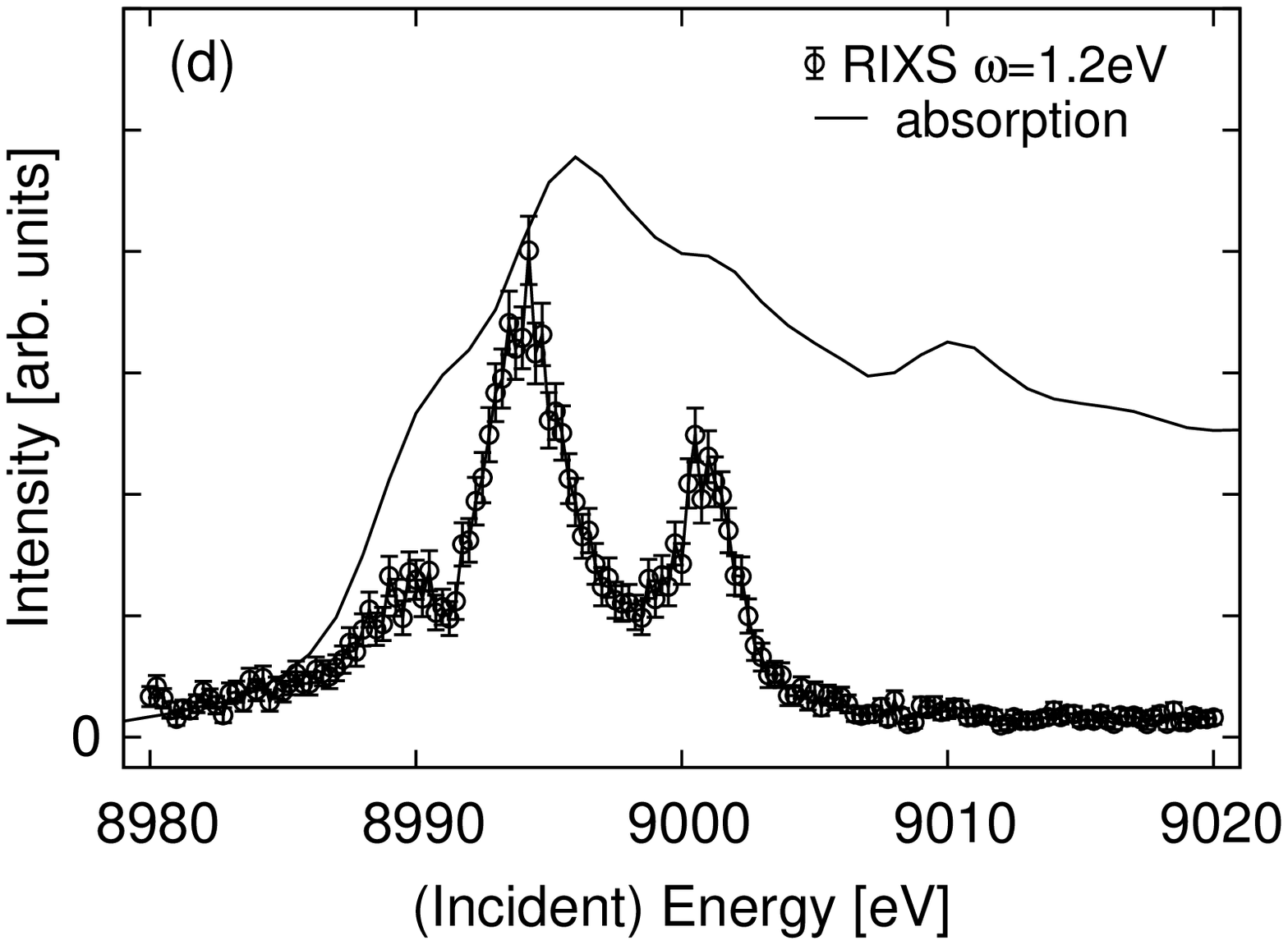}
\caption{(Color online) (a) Schematic representation of the orbital
order of KCuF$_3$ in the hole picture.  (b) Energy levels of the $d$
electrons in KCuF$_3$ taken from $E_{\rm max}$ in
Ref~\cite{Deisenhofer2008a}.  Solid and dashed arrows indicate the $e_g$
excitation and $t_{2g}$ excitations, respectively.  (c) Typical RIXS
spectrum of KCuF$_3$.  The vertical arrows indicate the peaks discussed
in the text.  (d) X-ray absorption spectrum of polycrystalline KCuF$_3$
(solid line) and RIXS intensity at 1.2-eV energy loss as a function of
the incident energy.}
\label{fig:eidep}
\end{figure*}

In this Letter, we conduct polarization analysis of the Cu $K$-edge RIXS
of the orbitally-ordered compound KCuF$_3$ and demonstrate that the
observation of the orbital ($d$-$d$) excitations by RIXS depends on the
polarization conditions.  KCuF$_3$ has long been known to display quantum
one-dimensional antiferromagnetic properties along the $c$-axis
originating from the strong super-exchange interaction between the $e_g$
orbitals of Cu$^{2+}$ \cite{Hirakawa1967}.  While the two $e_g$ orbitals
are degenerate under the cubic symmetry, in the real tetragonal
structure the degeneracy is lifted, accompanied with the occurrence of
strong Jahn-Teller distortions.  The $e_g$ orbitals are ordered
according to the pattern shown in Fig.~\ref{fig:eidep}(a), which
corresponds to an alternation of the hole orbitals $3d_{y^2-z^2}$ and
$3d_{z^2-x^2}$ on adjacent Cu sites.  There are two types of orbital
excitations in KCuF$_3$: One excitation is a transition of an electron
from the $t_{2g}$ orbital to the $e_g$ orbital and the other one is
between the $e_g$ orbitals.  The latter excitation is unique to the
orbital-ordered Mott insulators.  Utilizing our new
polarization-analyzed RIXS technique, we have successfully measured the
polarization dependence of these two types of orbital excitations and
found that they are distinct from each other.  The differences are
interpreted based on a phenomenological group-theoretical consideration
of RIXS, which gives the necessary polarization conditions for observing
each excitation.  In the last part, we show that the orbital excitations
are dispersionless within our experimental resolution.

The experiments were carried out at BL11XU at SPring-8.  Incident x-rays
were monochromatized by a double crystal Si(111) monochromator and a
secondary Si(400) channel-cut monochromator.  Scattered x-rays were
analyzed in energy by a spherically bent Ge(733) crystal.  When we
performed polarization analysis of the scattered x-rays, a Ge(800)
crystal was used instead of the Ge(733) one to keep space in the
spectrometer for the polarization-analyzing device.  The total energy
resolution estimated from the full width at half maximum of the elastic
peak was about 400 and 600 meV for the Ge(733) and Ge(800) crystals,
respectively.  For the polarization analysis of the scattered x-rays, a
pyrolytic graphite (PG) crystal was placed in front of the detector and
the (006) reflection of PG was measured.  The scattering angle
($\theta_{\rm P}$) of the reflection at 8994 eV is 38 $^{\circ}$ and the
polarization extinction ratio ($\sin^2 2\theta_{\rm P}$) is 0.94.
Experimental reflectivity of the (006) reflection was about 0.02.  By
rotating the PG crystal and the detector about the axis of the beam, one
can select the polarization of the scattered photon.

A single crystal of KCuF$_3$ was used.  Polytype structures
corresponding to different stacking of the $ab$-plane have been reported
to exist in KCuF$_3$ \cite{Okazaki1969}.  Our single crystal was
carefully prepared so that it only contains the (a)-type structure.
This was confirmed by a magnetic susceptibility measurement where we
observed a single antiferromagnetic transition at 38 K
\cite{Hutchings1969}.  We use the Miller indices in the tetragonal unit
cell of the primitive perovskite structure ($a$ = 4.1410 \AA\ and $c$ =
3.9237 \AA) to represent the momentum transfer (${\bm Q}$). The
propagation vector of the orbital order is $(1/2,1/2,1/2)$.  All the
spectra were measured at room temperature.

A typical RIXS spectrum of KCuF$_3$ is shown in Fig.~\ref{fig:eidep}(c).
The incident photon energy ($E_{\rm i}$) is 8994 eV and the momentum
transfer is ${\bm Q}=(0,0,3.9)$.  Three peaks are observed at around
1.2, 6.5, and 8.5 eV, as indicated by the arrows in
Fig.~\ref{fig:eidep}(c).  The 1.2-eV peak corresponds to orbital
excitations which are the subject of this letter.  A band structure
calculation in the local density approximation including the on-site
Coulomb interaction (LDA+U) suggested that KCuF$_3$ is a charge-transfer
insulator rather than a Mott insulator \cite{Binggeli2004}.  Therefore
we tentatively ascribe the 6.5-eV peak to a charge-transfer excitation
from F $2p$ to Cu $3d$.  It is noted that the calculated band gap is
smaller than the energy loss of the RIXS peak. Using the analogy with
copper oxides \cite{Kim2004b}, we are tempted to ascribe the 8.5-eV peak
to a molecular orbital excitation between bonding and antibonding states
in a CuF$_6$ octahedron.

In Fig.~\ref{fig:eidep}(d), we show the resonant profile of the 1.2-eV
excitation obtained by plotting the RIXS intensity at a fixed energy
transfer ($\omega$) of 1.2 eV as a function of $E_{\rm i}$.  The x-ray
absorption spectrum (XAS) measured on a polycrystalline sample is
superimposed.  The resonant enhancement is found to occur for incident
energies in the vicinity of the salient features of the XAS spectrum.
Hereafter the incident photon energy is fixed at 8994 eV, where the
intensity of the 1.2 eV excitation is maximum.

\begin{figure*}[t]
\includegraphics[scale=0.24]{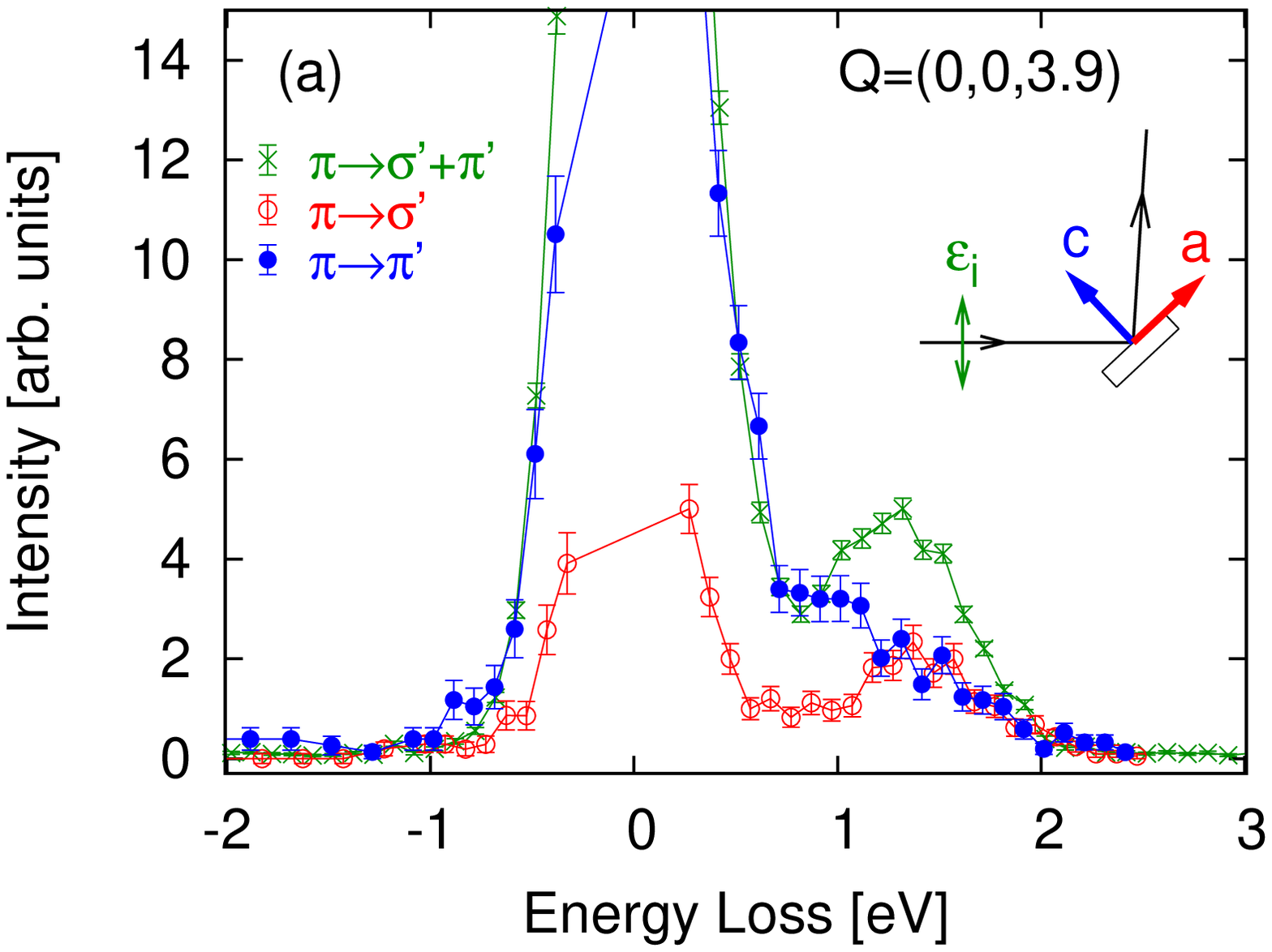}
\includegraphics[scale=0.24]{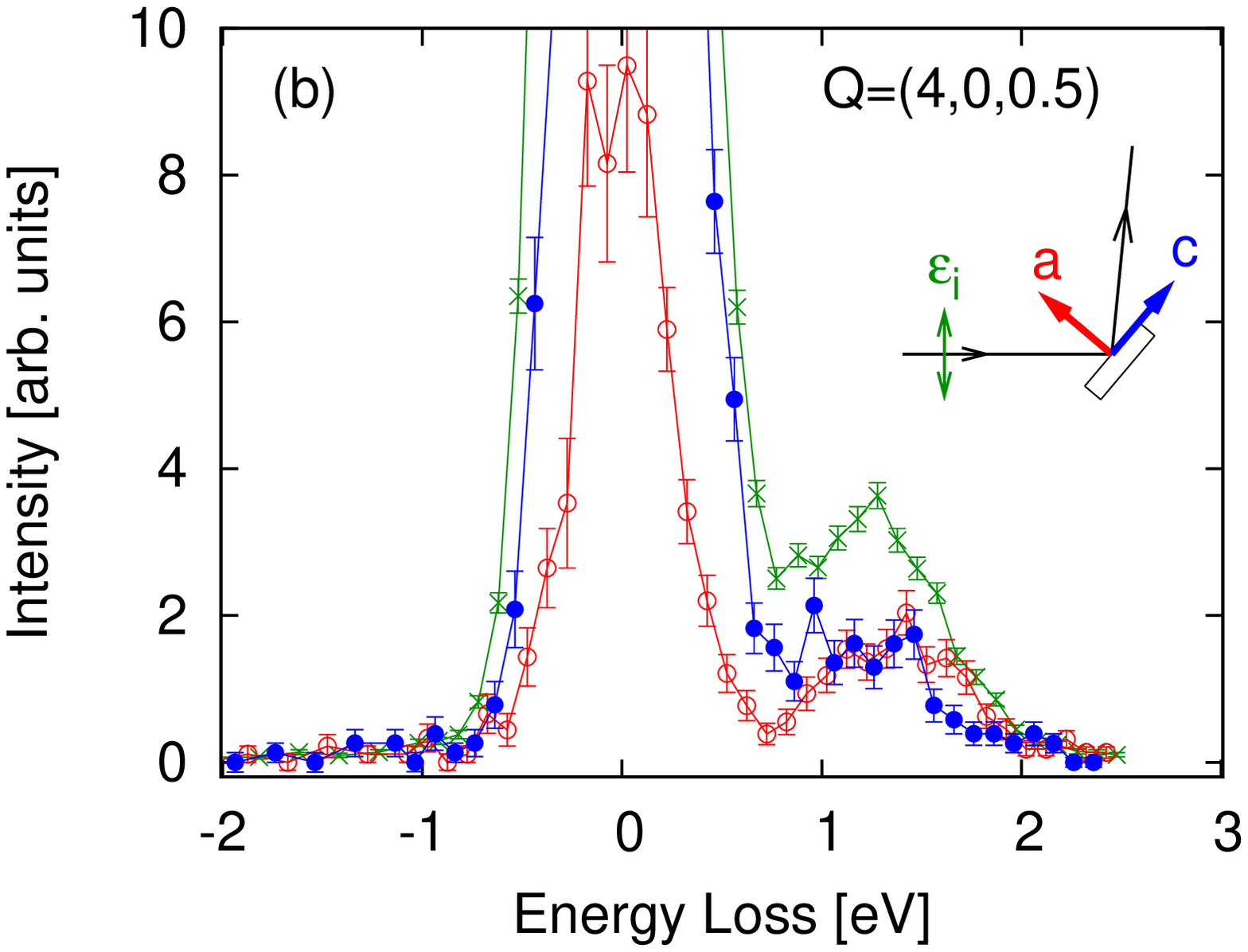}
\includegraphics[scale=0.24]{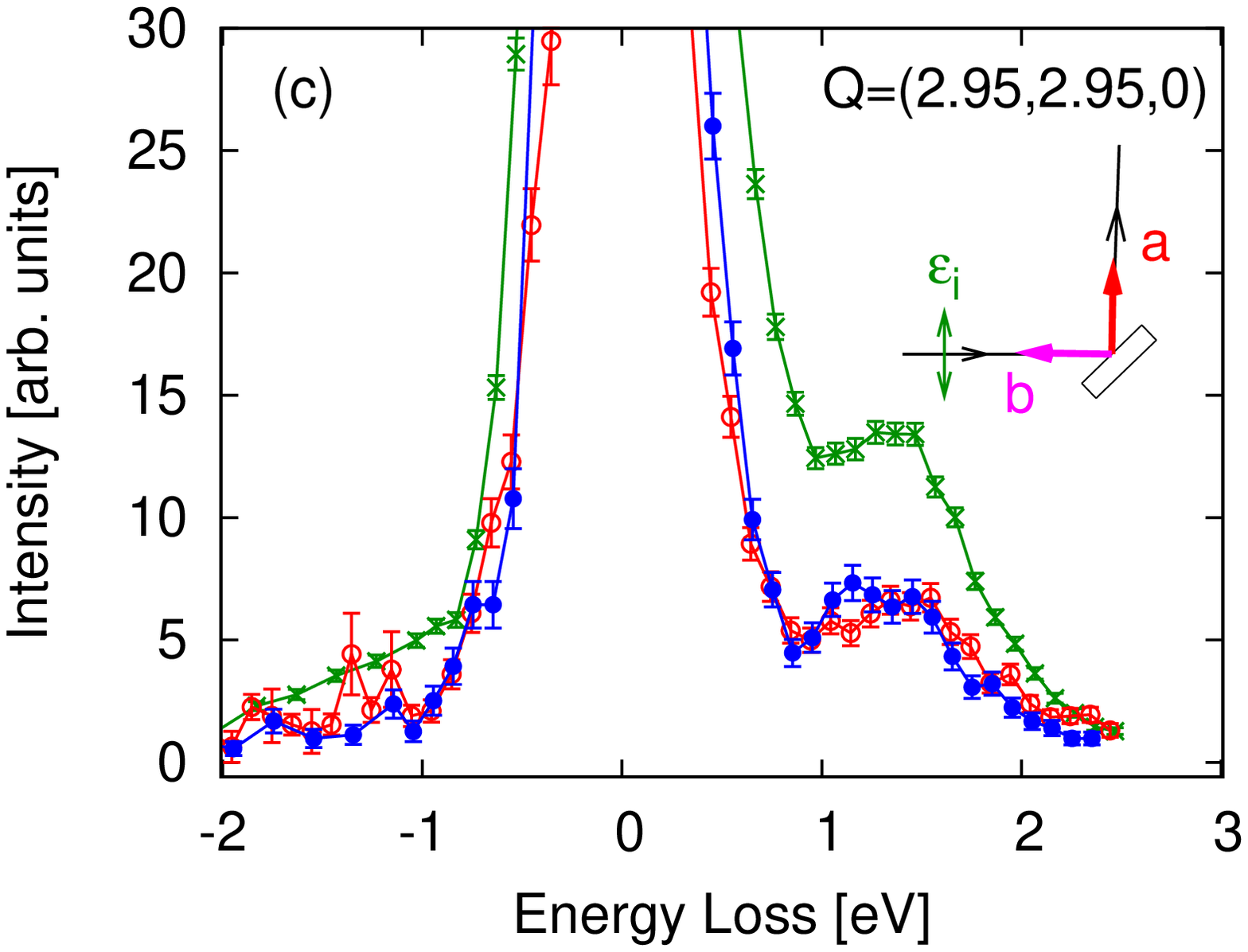}
\includegraphics[scale=0.24]{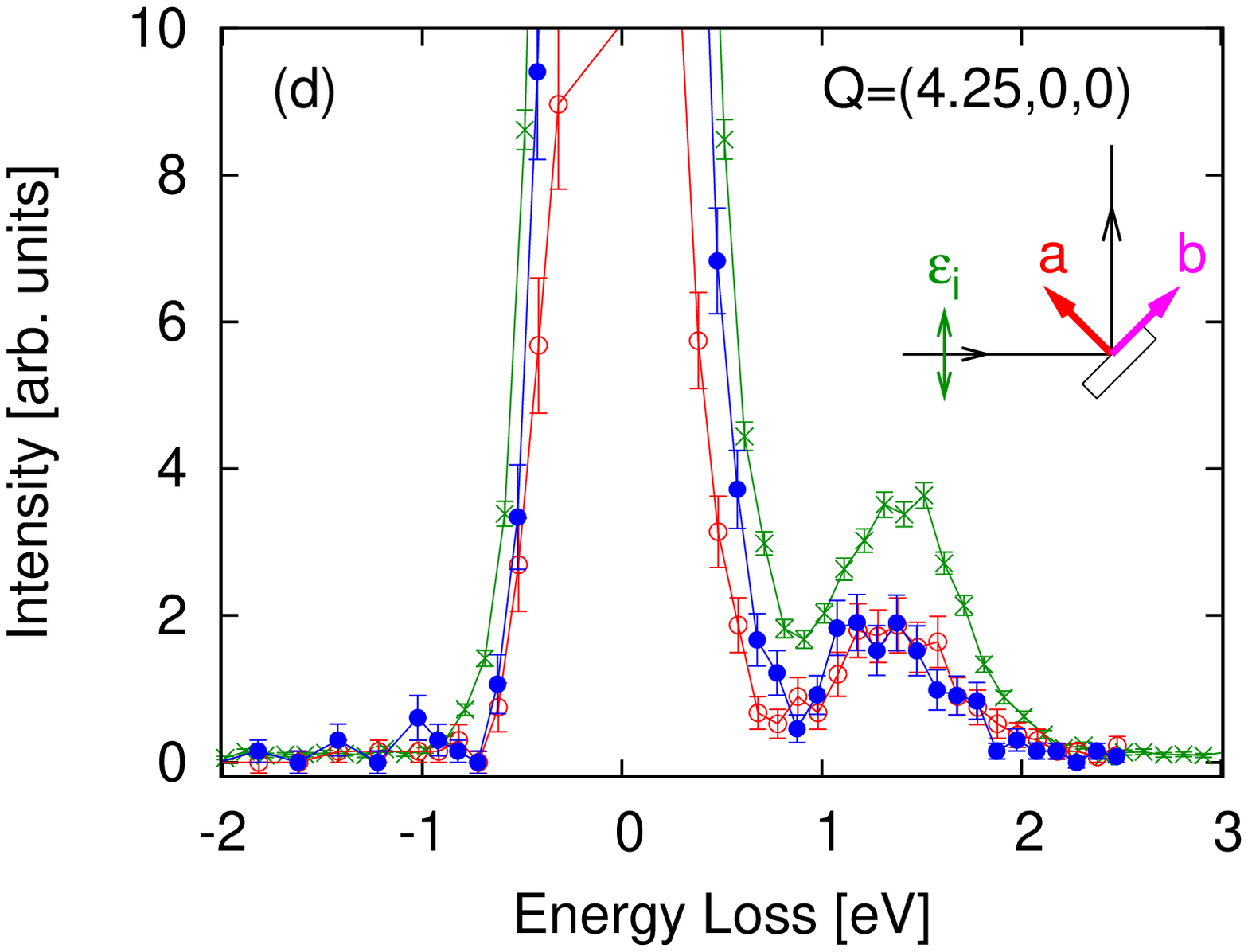}
\caption{(Color online) Polarization-analyzed RIXS spectra.  Spectra
with filled and open circles are measured in the $\pi\to\sigma'$ and
$\pi\to\pi'$ conditions, respectively. The spectra without polarization
analysis are also shown (crosses).  Corresponding Experimental
geometries are shown as inset.}
\label{fig:poldep}
\end{figure*}

The polarization-analyzed RIXS spectra of the 1.2-eV feature are
presented in Figs.~\ref{fig:poldep} for four different crystal
orientations schematized in the corresponding insets.  The incident
photon polarization (${\boldsymbol \epsilon}_{\rm i}$) lies in the
scattering plane and the scattering angle ($2\theta$) is chosen to be
close to 90$^{\circ}$ in order to reduce the elastic scattering.  The
incident and scattered polarizations are therefore orthogonal to each
other, which is the so-called the depolarized configuration.  In each plot,
the spectra are normalized so that the sum of the intensity of the
$\pi\to\sigma'$ and $\pi\to\pi'$ inelastic signals is equal to the
$\pi\to\sigma'+\pi'$ one. Here $\pi$ is the incident photon polarization
parallel to the scattering plane while $\pi'$ and $\sigma'$ denote the
scattered photon polarizations parallel and perpendicular to the
scattering plane, respectively.  The most striking observation of this
measurement is that additional intensity is found around 1 eV in the
spectrum of $\pi\to\pi'$ compared with $\pi\to\sigma'$ in the
configurations of (a) and (b), whereas the spectra in (c) and (d) are
almost identical between the two polarization conditions.  In a recent
optical absorption study \cite{Deisenhofer2008a,Deisenhofer2008anote},
the different orbital excitations were identified and their respective
energies were estimated.  The corresponding results are summarized in
Fig.~\ref{fig:eidep}(b).  Based on this assignment, we can ascribe the
additional intensity at 1 eV to the excitation between the $e_g$
orbitals ($e_g$ excitation).  Excitations from one of the $t_{2g}$
orbitals to the unoccupied $e_g$ orbital ($t_{2g}$ excitations) are
respectively located at 1.15, 1.37, and 1.46 eV in the optical
absorption spectrum.  These excitations are found to appear as a single
peak in the RIXS spectrum within our resolution.  While the $t_{2g}$
excitations are observed for both polarization conditions in all four
geometrical configurations, the $e_g$ excitation appears only for the
$\pi\to\pi'$ polarization of the configurations of
Figs.~\ref{fig:poldep} (a) and (b).  This is a direct evidence that the
$e_g$ excitation has a different polarization dependence from the
$t_{2g}$ excitations.

The polarization conditions corresponding to the different geometrical
configurations in Figs.~\ref{fig:poldep} are summarized in
Table~\ref{tab:pol}.  The unit lattice vectors of the crystal are noted
as ${\bm a}$, ${\bm b}$, and ${\bm c}$.  The unit vectors ${\bm x}$,
${\bm y}$, and ${\bm z}$ are taken so that the hole orbital is
represented as $x^2-y^2$.  It is noted that there are two polarization
conditions in the $xyz$ coordinates because of the two orbital
sublattices as illustrated in Fig.~\ref{fig:eidep}(a).

\begin{table}[b]
\caption{Summary of the polarization conditions of
Fig.~\ref{fig:poldep}.}  
\label{tab:pol}
\begin{tabular}{c|c|cc|c}
\hline \hline
\multirow{2}{*}{config.}&\multirow{2}{*}{polarization}&
\multirow{2}{*}{\makebox[4em]{${\bm \epsilon}_i$}}&
\multirow{2}{*}{\makebox[4em]{${\bm \epsilon}_f$}}&
symmetry of\\
&&&&$P_i \times P_f$\\
\hline
\multirow{4}{*}{(a)}
&$\pi\to\sigma'$
&${\bm y}+{\bm z}$&${\bm x}$&$A_{2g}+B_{2g}+E_g$\\
&$({\bm a}+{\bm c})\to{\bm b}$
&${\bm x}+{\bm y}$&${\bm z}$&$E_g$\\
\cline{2-5}
&$\pi\to\pi'$
&${\bm y}+{\bm z}$&${\bm y}-{\bm z}$&$A_{1g}+B_{1g}+E_g$\\
&$({\bm a}+{\bm c})\to({\bm a}-{\bm c})$
&${\bm x}+{\bm y}$&${\bm x}-{\bm y}$&$B_{1g}+A_{2g}$\\
\hline
\multirow{4}{*}{(b)}
&{$\pi\to\sigma'$}
&${\bm y}+{\bm z}$&${\bm x}$&$A_{2g}+B_{2g}+E_g$\\
&$({\bm a}+{\bm c})\to{\bm b}$
&${\bm x}+{\bm y}$&${\bm z}$&$E_g$\\
\cline{2-5}
&{$\pi\to\pi'$}
&${\bm y}+{\bm z}$&${\bm y}-{\bm z}$&$A_{1g}+B_{1g}+E_g$\\
&$({\bm a}+{\bm c})\to({\bm a}-{\bm c})$
&${\bm x}+{\bm y}$&${\bm x}-{\bm y}$&$B_{1g}+A_{2g}$\\
\hline
\multirow{4}{*}{(c)}
&$\pi\to\sigma'$
&${\bm z}$&${\bm y}$&$E_g$\\
&${\bm a}\to{\bm c}$
&${\bm y}$&${\bm x}$&$A_{2g}+B_{2g}$\\
\cline{2-5}
&$\pi\to\pi'$
&${\bm z}$&${\bm x}$&$E_g$\\
&${\bm a}\to{\bm b}$
&${\bm y}$&${\bm z}$&$E_g$\\
\hline
\multirow{4}{*}{(d)}
&$\pi\to\sigma'$
&${\bm x}+{\bm z}$&${\bm y}$&$A_{2g}+B_{2g}+E_g$\\
&$({\bm a}+{\bm b})\to{\bm c}$
&${\bm y}+{\bm z}$&${\bm x}$&$A_{2g}+B_{2g}+E_g$\\
\cline{2-5}
&$\pi\to\pi'$
&${\bm x}+{\bm z}$&${\bm x}-{\bm z}$&$A_{1g}+B_{1g}+E_g$\\
&$({\bm a}+{\bm b})\to({\bm a}-{\bm b})$
&${\bm y}+{\bm z}$&${\bm y}-{\bm z}$&$A_{1g}+B_{1g}+E_g$\\
\hline \hline
\end{tabular}
\end{table}

In order to interpret the difference between the polarization dependence
of the $e_g$ excitation and the $t_{2g}$ excitations, we treat
theoretically the polarization dependence of RIXS from a
phenomenological view point, without identifying the microscopic
scattering processes.  The scattering cross section in RIXS is
represented by the correlation function of the polarizability operator,
corresponding to the $S$-matrix, as a function of momentum and the
photon polarizations \cite{Ishihara2000a,Ishihara2004}.  This operator
is expanded in terms of the excitation modes based on the group
theoretical analyses.  When the electronic excitations occur at the
local site where the x-ray is absorbed, there is a selection rule for
RIXS that a product-representation $\Gamma_i \times \Gamma_f \times P_i
\times P_f$ should belong to the $A_{1g}$ symmetry.  Here
$\Gamma_{i(f)}$ and $P_{i(f)}$ are the irreducible representations for
the initial (final) electronic orbital and those of the incident
(scattered) photon polarization, respectively.  Equivalently, the
excitations are allowed only if the product-representations $\Gamma_i
\times \Gamma_f$ and $P_i \times P_f$ share at least one common
symmetry.  When other sites are involved in the electronic excitations,
this selection rule is modified and depends on the momentum transfer
${\bf Q}$.

\begin{table}[b]
\caption{Symmetry of the orbital excitations of KCuF$_3$.}
\label{tab:excitation}
\begin{tabular}{c|c}
\hline \hline
excitation&symmetry of $\Gamma_i\times\Gamma_f$\\
\hline
$(3z^2-r2) \to (x^2-y^2)$&$B_{\rm 1g}$\\
$(xy) \to (x^2-y^2)$&$A_{\rm 2g}$\\
$(yz) \to (x^2-y^2)$&$E_{\rm g}$\\
$(zx) \to (x^2-y^2)$&$E_{\rm g}$\\
\hline \hline
\end{tabular}
\end{table}

We approximately treat the local symmetry of the Cu atom as $D_{4h}$
instead of the exact site symmetry of $D_{2h}$ because the Cu-F
distances along the $x$ and $y$ directions are almost equal.  If only
considering on-site excitations, the possible excitation modes in each
polarization configuration of the measurements are given by $P_i \times
P_f$. Those are listed in Table~\ref{tab:pol}.  We also show the
symmetry of the orbital excitations $\Gamma_i \times \Gamma_f$ in
Table~\ref{tab:excitation}.  The symmetry of $P_i \times P_f$ has to be
either $A_{\rm 2g}$ or $E_{\rm g}$ for the observation of the $t_{2g}$
excitation and either symmetry is always present in all four
experimental configurations of Figs.~\ref{fig:poldep}.  It means that
the theoretical prediction agrees with the experimental results for the
$t_{2g}$ excitation.  On the other hand, $e_g$ excitation requires
$B_{\rm 1g}$ polarization symmetry, which exists in the $\pi\to\pi'$
condition for the configurations of Figs.~\ref{fig:poldep} (a), (b), and
(d).  Experimentally, the $e_g$ excitation is observed in the
$\pi\to\pi'$ condition for the configurations of Figs.~\ref{fig:poldep}
(a) and (b).  The symmetry argument is again valid for the $e_g$
excitation as long as it gives a necessary condition for its observation
by RIXS.  We briefly comment on the absence of the $e_g$ excitation in
the spectrum obtained in the $\pi\to\pi'$ condition of the configuration
for Fig.~\ref{fig:poldep} (d). From our symmetry analysis, the weight of
the $B_{1g}$ component for the $\pi\to\pi'$ polarization in the
configuration of Fig.~\ref{fig:poldep} (d) is deduced to be 40\% of that
for the $\pi\to\pi'$ polarization of Figs.~\ref{fig:poldep} (a) and (b).
For a quantitative agreement with the experiment, one may need to carry
out a theoretical evaluation of the intensity including a microscopic
description of the RIXS process.

\begin{figure}[t]
\includegraphics[scale=0.24]{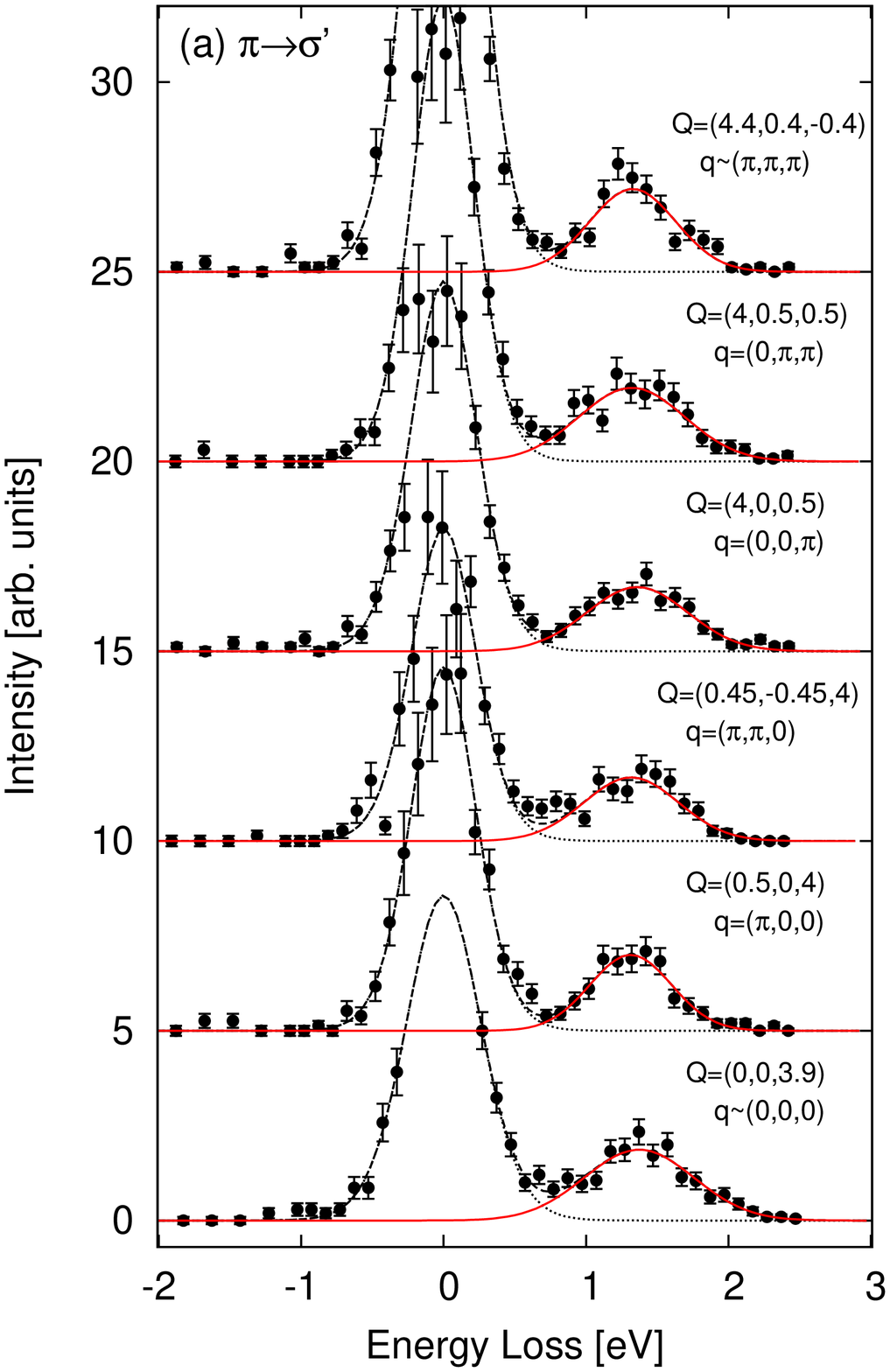}
\includegraphics[scale=0.24]{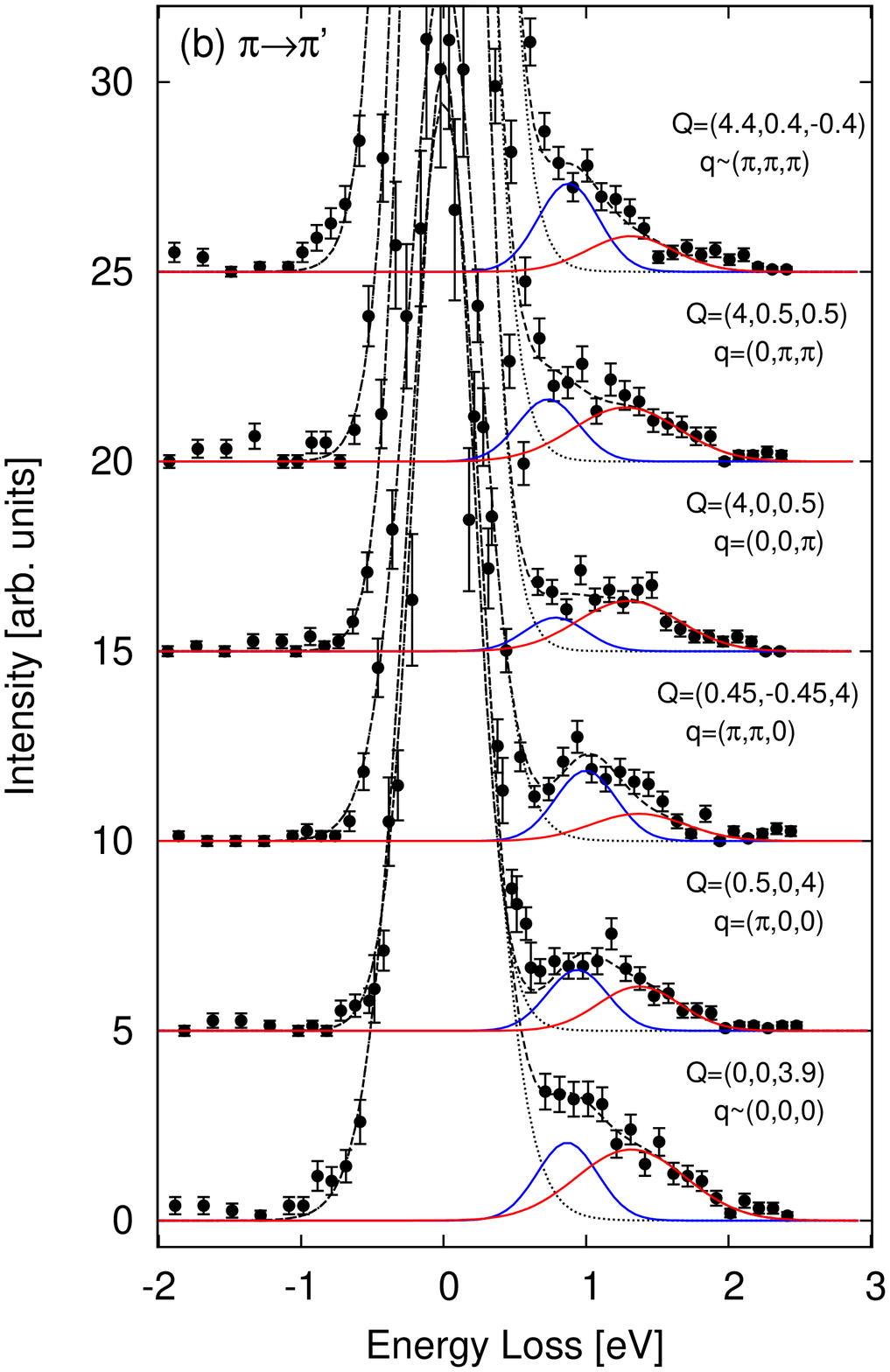}
\includegraphics[scale=0.33]{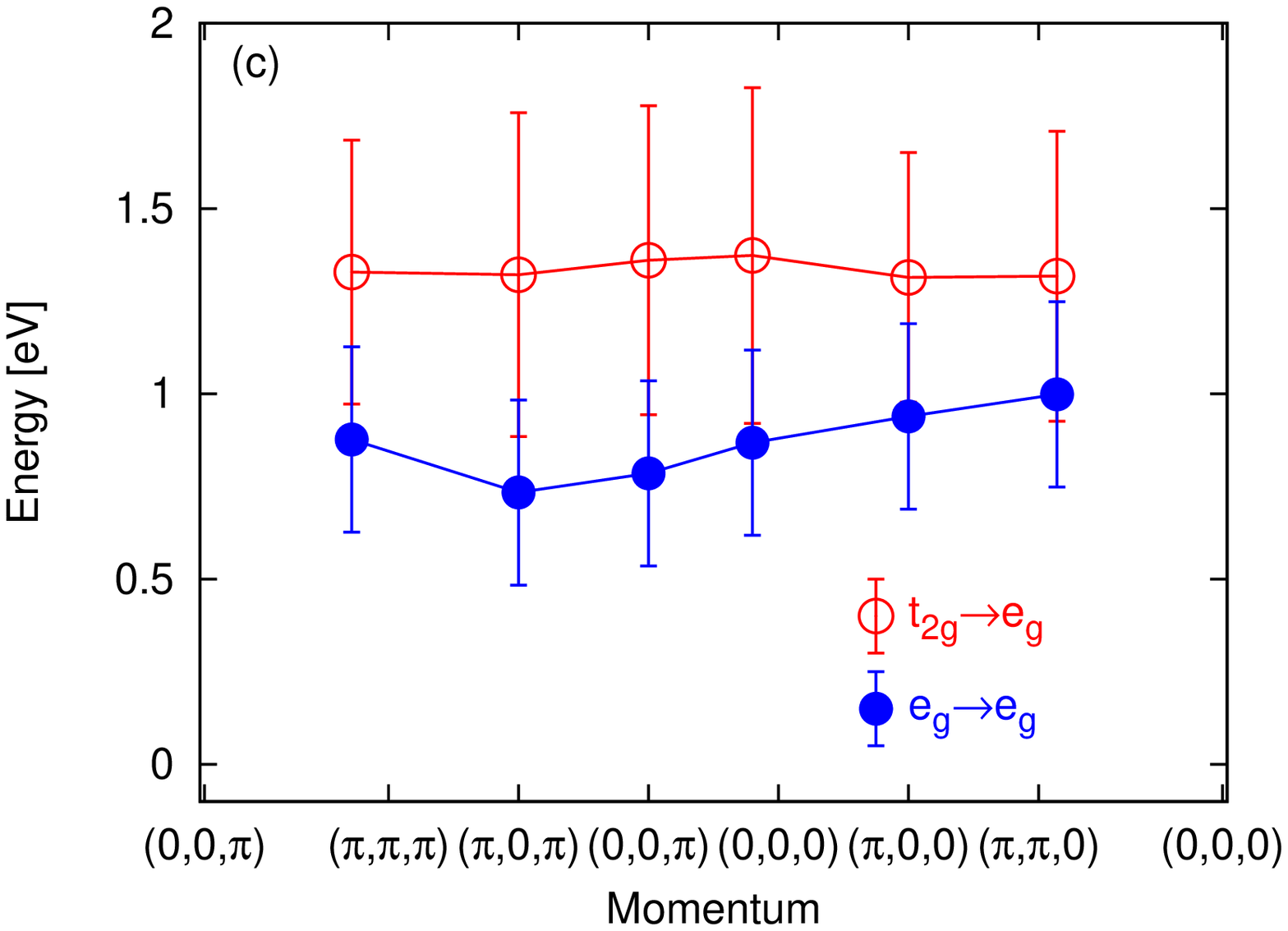}
\caption{(Color online) (a) and (b) Momentum dependence of
polarization-analyzed RIXS spectra.  ${\bm Q}$ and ${\bm q}$ are the
absolute momentum transfer and the reduced one, respectively.  The
filled circles correspond to the experimental data and solid-line
spectra are the fitted orbital excitations. Dashed-line spectra
represent the sum of the fitted elastic line (dotted line) with the
fitted orbital excitations (solid line).  (c) Dispersion relation of the
orbital excitations.  The peak positions are indicated by the circles
and the bars denote the FWHM of the excitations.}
\label{fig:qdep}
\end{figure}

We now turn to the momentum dependence of the orbital excitations.
Because the orbital is an important degree of freedom through which
novel properties of transition metal compounds can be controlled, the
understanding of the elementary orbital excitation, so-called orbiton,
is still a fascinating issue.  Using Raman scattering, orbital
excitation at zero momentum transfer has been observed in LaMnO$_3$
\cite{Saitoh2001}.  RIXS makes it feasible to probe the momentum
dependence of this excitation \cite{Ishihara2000}.  Owing to our
polarization analysis, we have been able to undoubtedly assign the 1 eV
feature in the RIXS spectrum of KCuF$_3$ to the orbital degree of
freedom of the system.  We now present the momentum dependence of this
excitation obtained for the whole Brillouin zone.  Figures
\ref{fig:qdep} (a) and (b) show the polarization-analyzed RIXS spectra
obtained near the high-symmetry points of the Brillouin zone.  Here
${\bm q}$ is the momentum transfer in the reduced Brillouin zone.  The
spectra of the upper and lower three momenta in each figure are measured
at or near the configurations of Fig.~\ref{fig:poldep} (a) and (b),
respectively.

In order to elucidate the dispersion relation of the orbital excitations
quantitatively, we fitted the whole set of RIXS spectra in
Figs.~\ref{fig:qdep} (a) and (b).  We start with the spectra collected
with the $\pi\to\sigma'$ polarization in Fig.~\ref{fig:qdep} (a), where
only the $t_{2g}$ excitation is observed.  The tail of the elastic
scattering or quasielastic component on the energy loss side was
evaluated from the energy gain side.  The excitation peak was
approximated to a single Gauss function.  We then fitted the spectra
obtained in the $\pi\to\pi'$ condition in Fig.~\ref{fig:qdep} (b), using
the peak positions and widths of the $t_{2g}$ excitation inferred from
the fits of the spectra in Fig.~\ref{fig:qdep} (a) and keeping them
fixed.  The peak centers and FWHMs used in the fits are plotted as
functions of the reduced momentum transfer in Fig.~\ref{fig:qdep} (c).
The dispersion relation of the excitations is small.  It is noteworthy
that the dispersion of this excitation between the two $e_g$ orbitals,
which corresponds to the orbital order in KCuF$_3$, is not discernible
within the experimental resolution.  This result suggests that the
Jahn-Teller splitting between the two $e_g$ orbitals is about 1 eV
irrespective of the momentum transfer.  Moreover, we conclude that the
inter-site interactions between the $d$ orbitals, which can be ascribed
to the magnitude of the dispersion, are much smaller in energy than the
experimental resolution (600 meV).  Overall, these data importantly show
that RIXS at the $K$-edge can be used as a robust technique that
provides information about the dependence of the electronic excitations
upon two key parameters of the condensed matter physics, momentum and
polarization.  Since improvement of the energy resolution is sill in
progress, the dispersion relation of orbital excitations should become
observable by $K$-edge RIXS in the near future.

In summary, we have performed a polarization-analyzed resonant inelastic
x-ray scattering study of the orbital excitations in KCuF$_3$ at the Cu
$K$-edge.  The polarization of the scattered photons in RIXS was
identified for the first time.  A clear contrast is found in the
polarization dependence between the excitation between the $e_g$
orbitals and the excitations from the $t_{2g}$ orbitals to the $e_g$
orbital.  The former excitation is observed in the $\pi\to\pi'$
polarization for two of the four studied geometrical configurations,
while the latter appears in both $\pi\to\pi'$ and $\pi\to\sigma'$
polarizations for all four configurations.  The polarization dependence
of both orbital excitations can be understood based on the symmetry of
the RIXS process, which gives a necessary condition for the appearance
of these orbital excitations.  We measured the momentum dependence of
the polarization-analyzed excitations and found that both orbital
excitations are nearly dispersionless within our experimental
resolution.

This work was performed under the inter-university cooperative research
program of the Institute of Materials Research, Tohoku University and
financially supported by the Grant-in-Aid for Young Scientists (B)
(No.~20740179) from JSPS.

\bibliography{/home/kenji/tex/jabref/bib/rixs-experiment.bib,/home/kenji/tex/jabref/bib/ixs-theory.bib,/home/kenji/tex/jabref/bib/kcuf3.bib,/home/kenji/tex/jabref/bib/kcuf3-paper.bib,/home/kenji/tex/jabref/bib/soft-rixs.bib}

\begin{thebibliography}{27}
\bibitem[{Dei()}]{Deisenhofer2008anote}
\bibinfo{note}{In general, electric dipole transtions between $d$ orbitals are
  parity-forbidden but they can become allowed by assistance of phonons.}
\end{thebibliography}

\end{document}